 \documentclass[final,3p,times]{elsarticle}

\usepackage{epsfig}

\usepackage{mathtools}

\usepackage[english]{babel}
\usepackage[utf8]{inputenc}
\usepackage{subcaption}

\usepackage[hidelinks]{hyperref}
\hypersetup{colorlinks=true, urlcolor=blue, linkcolor=blue, citecolor=red}

\journal{Physica A: Statistical Mechanics and its Applications}

\begin{document}

\begin{frontmatter}

\title{A Further Analysis of The Role of Heterogeneity \\ in Coevolutionary Spatial Games \tnoteref{t1}}

\tnotetext[t1]{Published at \url{https://doi.org/10.1016/j.physa.2017.10.035}}

\author{Marcos Cardinot}
\ead{marcos.cardinot@nuigalway.ie}

\author{Josephine Griffith}

\author{Colm O'Riordan}

\address{Department of Information Technology, National University of Ireland, Galway, Ireland}

\begin{abstract}

Heterogeneity has been studied as one of the most common explanations of the
puzzle of cooperation in social dilemmas. A large number of papers have
been published discussing the effects of increasing heterogeneity in
structured populations of agents, where it has been established that
heterogeneity may favour cooperative behaviour if it supports agents
to locally coordinate their strategies. In this paper, assuming
an existing model of a heterogeneous weighted network, we aim to further
this analysis by exploring the relationship (if any) between heterogeneity
and cooperation. We adopt a weighted network which is fully populated by
agents playing both the Prisoner's Dilemma or the Optional Prisoner's
Dilemma games with coevolutionary rules, i.e., not only the strategies but
also the link weights evolve over time.  Surprisingly, results show that
the heterogeneity of link weights (states) on their own does not always
promote cooperation; rather cooperation is actually favoured by the
increase in the number of overlapping states and not by the heterogeneity
itself. We believe that these results can guide further research towards a
more accurate analysis of the role of heterogeneity in social dilemmas.

\end{abstract}


\end{frontmatter}


\section{Introduction}

Issues regarding the emergence of cooperation and altruism in structured
populations have puzzled scientists in a large range of domains. In this
context, methods of Statistical Physics combined with concepts of both Graph
Theory and Evolutionary Game Theory \cite{Nowak2006,Smith1982} have been used
as simple and powerful tools to describe and analyse the conflict of interest
between individuals and groups~\cite{Perc2017}. In those models, agents are
arranged on graphs in such a way that their interactions are restricted to
their immediate neighbours~\cite{Nowak2009,Lieberman2005}. Over the last two
decades, it has been shown that different topologies such as
lattices~\cite{Nowak1992}, scale-free
graphs~\cite{Szolnoki2016,Xia2015,Santos2005}, small-world
graphs~\cite{Chen2008,Fu2007}, cycle graphs~\cite{Altrock2017}, star-like
graphs~\cite{Teng2014} and bipartite graphs~\cite{Pena2012,Gomez2011} have a
considerable impact on the evolution of cooperation, which also favours the
formation of different patterns and phenomena~\cite{Szabo2007,Perc2013}.
However, the vast majority of these studies adopt static networks, which are
not suitable for modelling scenarios in which both the game strategies and the
network itself are subject to evolution
\cite{Zhang2011a,Zhang2011b,Hai2011,Szolnoki2009,Zimmermann2004,Zimmermann2001}.
Thus, the use of dynamic networks represents a natural upgrade of the
traditional spatial games~\cite{Perc2010}.

The Prisoner's Dilemma (PD) is still the most often used game in this field. In
this game, an agent can either cooperate (C) or defect (D), obtaining a payoff
that depends on the other's agent choice~\cite{Rapoport1965}. However, in many
scenarios, agents have the freedom to decide whether to participate in the game.
Games such as the Optional Prisoner's Dilemma
(OPD)~\cite{Szabo2002,Batali1995} and the Voluntary Public
Goods game~\cite{Lu2017,Hauert2008} incorporate this concept of voluntary
participation by adding a third strategy to the game, allowing agents to not
only cooperate or defect but also to abstain (A) from a game interaction.
Research has shown that the presence of abstainers in the population can
actually protect cooperators against
exploitation~\cite{CardinotSAB,Hauert2003}.

Studies on weighted networks have attracted much attention as such networks enable the
representation of the strength of each connection, which is essential
information in a wide range of real-world scenarios including biological
networks and social media.
Recently, both the Prisoner's Dilemma and the Optional Prisoner's Dilemma games
have been explored in the context of dynamic weighted networks, which lead to a
coevolutionary scenario where not only the game strategies, but also the link
weights, evolve over time \cite{CardinotSCI,CardinotAICS,Huang2015,Cao2011,Li2017}.
Moreover, it has been shown that the use of dynamic weighted networks can increase
heterogeneity of states (i.e., the number of possible utilities in the network),
which in turn induces the promotion of cooperation.
In fact, previous work has also discussed the effects of heterogeneity on the
evolution of cooperation
\cite{Fan2017,Iwata2016,Amaral2016,Perc2011,Wardil2011,Szolnoki2008a,Szolnoki2008b,Szolnoki2007,Santos2005},
however, the specific conditions that increase the diversity of link weights in
the dynamic weighted networks remain unclear. Also, a number of questions
regarding the evolutionary dynamics of the network itself remain to be
answered, such as:
\begin{itemize}
    \item How the link weights between agents evolve over time?

    \item How two parameters of the model ($\Delta$ and $\delta$) affect the
        link weight variance?

    \item Is there an optimum value of the two parameters $\Delta$ and
        $\delta$?

    \item Why higher values of $\delta$ promote cooperation best?

    \item Why the Coevolutionary Optional Prisoner's Dilemma game performs
        better than the Coevolutionary Prisoner's Dilemma game in adverse
        scenarios?

    \item Does the value of $\Delta$ affect the convergence speed in scenarios
        of full dominance of cooperation?
\end{itemize}

Thus, this work aims to answer these questions by analysing the micro-macro
behaviour of a population of agents playing both the Coevolutionary Prisoner's
Dilemma (CPD) and the Coevolutionary Optional Prisoner's Dilemma (COPD) game,
i.e., the classical PD and OPD games in a dynamic weighted network.
The remainder of this paper is organized as follows. Section~\ref{sec:methods}
describes the Monte Carlo simulation and the coevolutionary games adopted.
Section~\ref{sec:results} features the results. Finally,
Section~\ref{sec:conclusion} summarizes our findings and outlines future
work.

\section{Methodology}
\label{sec:methods}

This work adopts a weighted lattice grid with periodic conditions (i.e., a
toroid) fully populated with $N=102\times102$ agents playing a coevolutionary
game. Each agent on site $x$ interacts only with its eight immediate neighbours
(i.e., $k=8$, Moore neighbourhood). Both the Coevolutionary Prisoner's Dilemma
(CPD) game~\cite{Huang2015} and the Coevolutionary Optional Prisoner's
Dilemma (COPD) game~\cite{CardinotSCI} are considered.

Initially, each edge linking agents has the same weight $w=1$, which will
adaptively change according to their interaction. Also, each agent ($x$) is
initially assigned to a strategy with equal probability. For the CPD game,
each agent can be designated either as a cooperator ($s_x=C$) or defector
($s_x=D$), while in the COPD game, agents can also be designated as abstainer
($s_x=A$). Thus, strategies ($s_x={C,D,A}$) can be denoted by a unit vector
respectively as follows:
\begin{equation}
    \label{eq:strategies}
    C=\begin{pmatrix}1\\0\\0\end{pmatrix}, \ 
    D=\begin{pmatrix}0\\1\\0\end{pmatrix}, \ 
    A=\begin{pmatrix}0\\0\\1\end{pmatrix}.
\end{equation}

The games are characterized by the payoff obtained according to the pairwise
interaction of agent $x$ and its neighbour $y$. Accordingly, the agent $x$ may
receive a reward $\pi_{xy}(C,C)=R$ for mutual cooperation; a punishment
$\pi_{xy}(D,D)=P$ for mutual defection; $\pi_{xy}(D,C)=T$ for successful
defection (i.e., there is a temptation to defect); $\pi_{xy}(C,D)=S$ for
unsuccessful cooperation (well-known as the sucker's payoff); or the loner's
payoff ($L$), which is obtained when one or both agents abstain (i.e.,
$\pi_{xy}(A,C)=\pi_{xy}(A,D)=\pi_{xy}(C,A)=\pi_{xy}(D,A)=\pi_{xy}(A,A)=L$). We
adopt a weak version of both games, where the payoff $R=1$, $T=b$ ($1<b<2$),
$L=l$ ($0<l<1$) and $S=P=0$ without destroying the nature of the dilemma
\cite{Nowak1992}. Thus, the payoff matrix $\pmb{\pi}$ is given by:
\begin{equation}
    \label{eq:payoff}
    \pmb{\pi}=
    \begin{pmatrix}
        1 & 0 & l \\
        b & 0 & l \\
        l & l & l \\
    \end{pmatrix},
\end{equation}
where:
\begin{equation}
    \pi_{xy}(s_x, s_y) = s_x^{T} \pmb{\pi} s_y.
\end{equation}
The utility $u_{xy}$ of agent $x$ with its neighbour $y$ is calculated as follows:
\begin{equation}
    \label{eq:utility}
    u_{xy} = w_{xy} \pi_{xy},
\end{equation}
where $w_{xy}$ represents the symmetric link weight of their interaction, i.e.,
$w_{xy}=w_{yx}$.

A number of Monte Carlo (MC) simulations are carried out to explore the
micro-macro behaviour of both the strategies and the weighted network itself.
Each MC simulation comprises the following elementary steps. First an agent
$x$ is randomly selected to play the coevolutionary game with its $k=8$
neighbours, obtaining an accumulated utility expressed as:
\begin{equation}
    U_x = \sum_{y \in \Omega_x}u_{xy},
\end{equation}
where $\Omega_x$ denotes the set of neighbours of the agent $x$.
Second, the agent $x$ updates all the link weights in $\Omega_x$ by comparing
each utility $u_{xy}$ with the average accumulated utility
(i.e., $\bar{U}_{x}=U_x/k$) as follows:
\begin{equation}
    \label{eq:bigdelta}
    w_{xy} =
    \begin{dcases*}
        w_{xy} + \Delta  & if $u_{xy} > \bar{U}_{x}$, \\
        w_{xy} - \Delta  & if $u_{xy} < \bar{U}_{x}$, \\
        w_{xy}           & otherwise,
    \end{dcases*}
\end{equation}
where $\Delta$ is a constant such that $0 \le \Delta \le \delta$. In line
with previous work \cite{CardinotSCI,Huang2015,Wang2014}, the link weight
is corrected to satisfy $1-\delta \le w_{xy} \le 1+\delta$, where $\delta$
($0 \le \delta < 1$) defines the weight heterogeneity. Note that when
$\Delta=0$ or $\delta=0$, the link weight remains constant (${w=1}$),
which decays in the classical scenario for static networks, i.e.,
only the strategies evolve.
Finally, the agent $x$ updates its strategy by comparing its current
accumulated utility $U_x$ (i.e., considering the updated weights) with
the accumulated utility of one randomly selected neighbour ($U_y$)
such that, if $U_y>U_x$ agent $x$ copies $s_y$ with a probability
proportional to the utility difference as follows:
\begin{equation}
    \label{eq:prob}
    p(s_x \leftarrow s_y) = \frac{U_y-U_x}{k(T-P)},
\end{equation}
otherwise, agent $x$ keeps its strategy for the next step.

In one Monte Carlo step (MCS), each agent is selected once on average, which
means that the number of inner steps in each MCS is equal to the population
size. Simulations are run for a sufficiently long thermalization time ($10^6$ MCS).
Furthermore, to alleviate the effect of randomness and to ensure proper accuracy in the
approach, the final results are obtained by averaging 10 independent runs.
It is noteworthy that due to the introduction of the weight factor ($w$) and
the quenched heterogeneities via $\delta$, the model is prone to evolve into
frozen patterns which represent quenched spatial randomness where a Griffiths
phase \cite{Griffiths1969} can emerge. As has been discussed in previous
studies \cite{Perc2011,Droz2009}, the evolutionary dynamics in these scenarios
tend to be very slow, which introduces some technical difficulties in classifying
the final stationary state. This is because the transition of the clusters of the
subordinate strategy into the dominant strategy requires that a large number
of the subordinate agents swap their strategies in a short period of time,
which is an occurrence that is very difficult.

\section{Results}
\label{sec:results}

In this section, we present some of the relevant experimental results obtained
when simulating a population of agents playing both the Prisoner's Dilemma and
the Optional Prisoner's Dilemma game on weighted networks.

\subsection{Exploring the coevolutionary rules}

As discussed in previous research \cite{CardinotSCI}, one interesting property
of the ratio $\Delta/\delta$ is that for any combination of both parameters
$\Delta$ and $\delta$, if their ratio is the same, then the number of states
is also the same. For instance, the pairs $(\Delta=0.02,\ \delta=0.2)$
and $(\Delta=0.08,\ \delta=0.8)$ have both $21$ possible link weights (states).

\begin{figure}[t]
\centering
{\epsfig{file=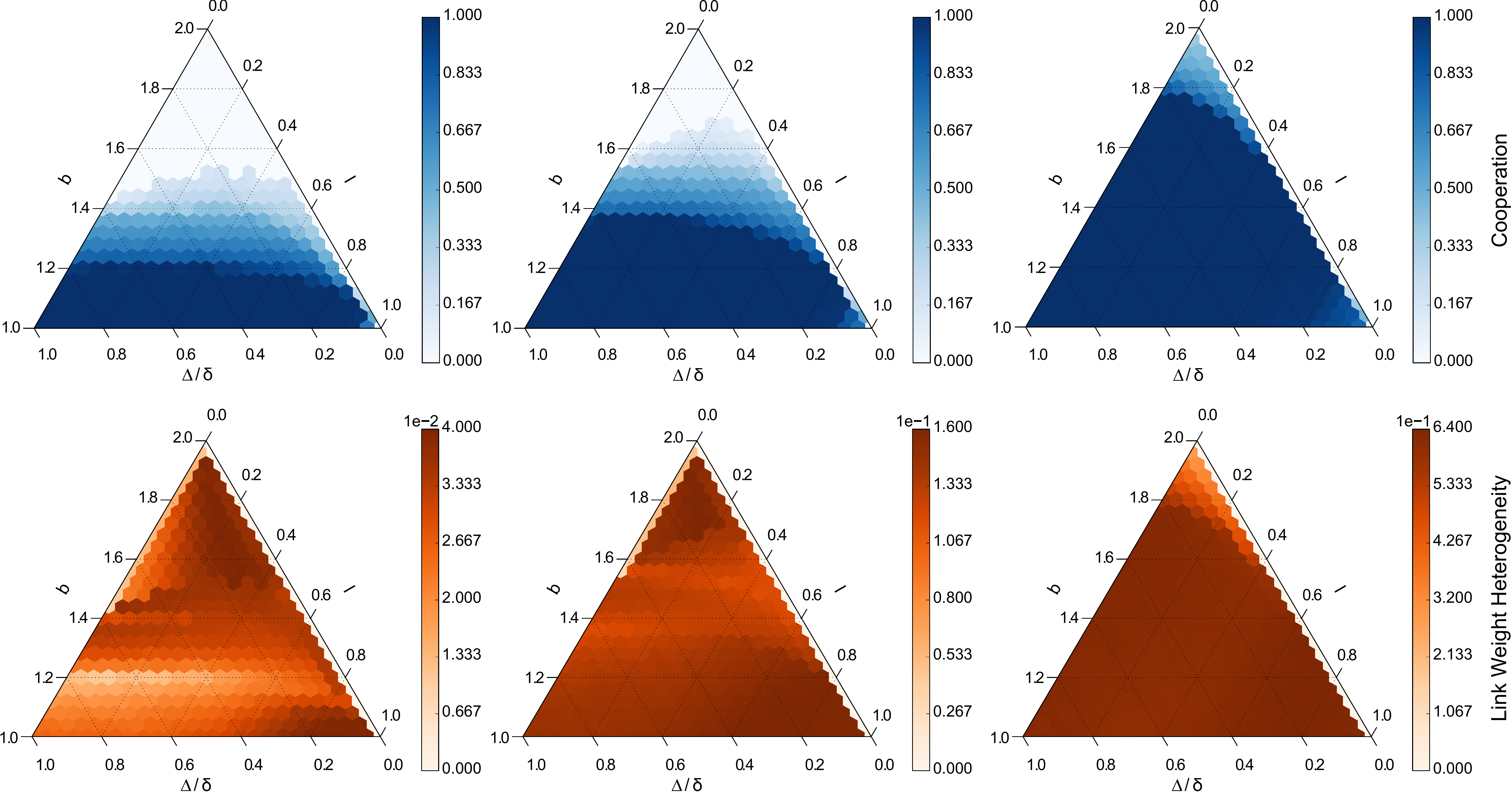, width=0.90\linewidth}}
\caption{
    Ternary diagrams showing the average fraction of cooperation (top) and the
    link weight heterogeneity (bottom) for different values of $b$ (i.e.,
    temptation to defect), $l$ (i.e., the loner's payoff) and $\Delta/\delta$
    for ${\delta=\{0.2,\ 0.4,\ 0.8\}}$ (from left to right).  The diagrams allow
    us to investigate the effects of different parameter settings for fixed
    values of $\delta$. The diagrams at the top suggest that higher values of
    $\delta$ may promote cooperation best, while the ones at the bottom show
    that the correlation between cooperation and heterogeneity does not seem to
    hold for all values of $\delta$.
}
\label{fig:ternary}
\end{figure}

Despite the fact that the interval between the maximum and minimum link weights
increases as we increase $\delta$, intuition may lead us to believe that given
two scenarios with the same payoff matrix (i.e., Eq.~\ref{eq:payoff} for the
same temptation to defect and loner's payoff) and the same number of
states (i.e., the same ratio $\Delta/\delta$), the outcome would be the same.
Surprisingly, previous research has shown that it is not true
\cite{CardinotSCI,Huang2015}. Actually, it has been discussed that higher
values of $\delta$ promote cooperation best, even if the number of states remains
the same.
Figure~\ref{fig:ternary} illustrates this scenario, where the average
fraction of cooperation of the last $10^3$ steps for a wider range
of settings (i.e., the ratio $\Delta/\delta$, the loner's payoff $l$ and
the temptation to defect $b$) are considered for different values of
$\delta$ (i.e., $\delta=\{0.2,\ 0.4,\ 0.8\}$).
In fact, the outcomes of the different scenarios are very different to each
other and we can observe that as $\delta$ increases, the number of cases in
which cooperation is the dominant strategy also increases.
This phenomenon is still unexplained and exploring the properties
that cause this discrepancy may lead to a complete understanding
of the presented coevolutionary model.
In any case, it is noteworthy that even for small values of $\delta$, the
Coevolutionary Optional Prisoner's Dilemma (COPD) game is still much more
beneficial for the emergence of cooperation than the traditional OPD
\cite{CardinotSAB} or the Coevolutionary Prisoner's Dilemma (CPD)
\cite{Huang2015}.

As is also shown in Figure~\ref{fig:ternary}, we investigate the influence
of the ratio $\Delta/\delta$ in the link weight variance (i.e., link weight
heterogeneity) at the last Monte Carlo steps. Note that as the link weight
($w$) is always within the range $[1-\delta,\ 1+\delta]$, the maximum link
weight variance is defined by $\delta^2$. These experiments reveal that the
link weight variance is not uniform for all environmental settings. Also,
although higher values of $\delta$ promote higher heterogeneity, we can see
that the correlation between cooperation and heterogeneity is not necessarily
true for all values of $\delta$. For instance, the link weight variance for
$\delta=0.8$ is usually maximum when cooperation is the dominant strategy.
However, this does not hold for both $\delta=0.4$ and $\delta=0.2$.

\begin{figure}[t]
\centering
{\epsfig{file=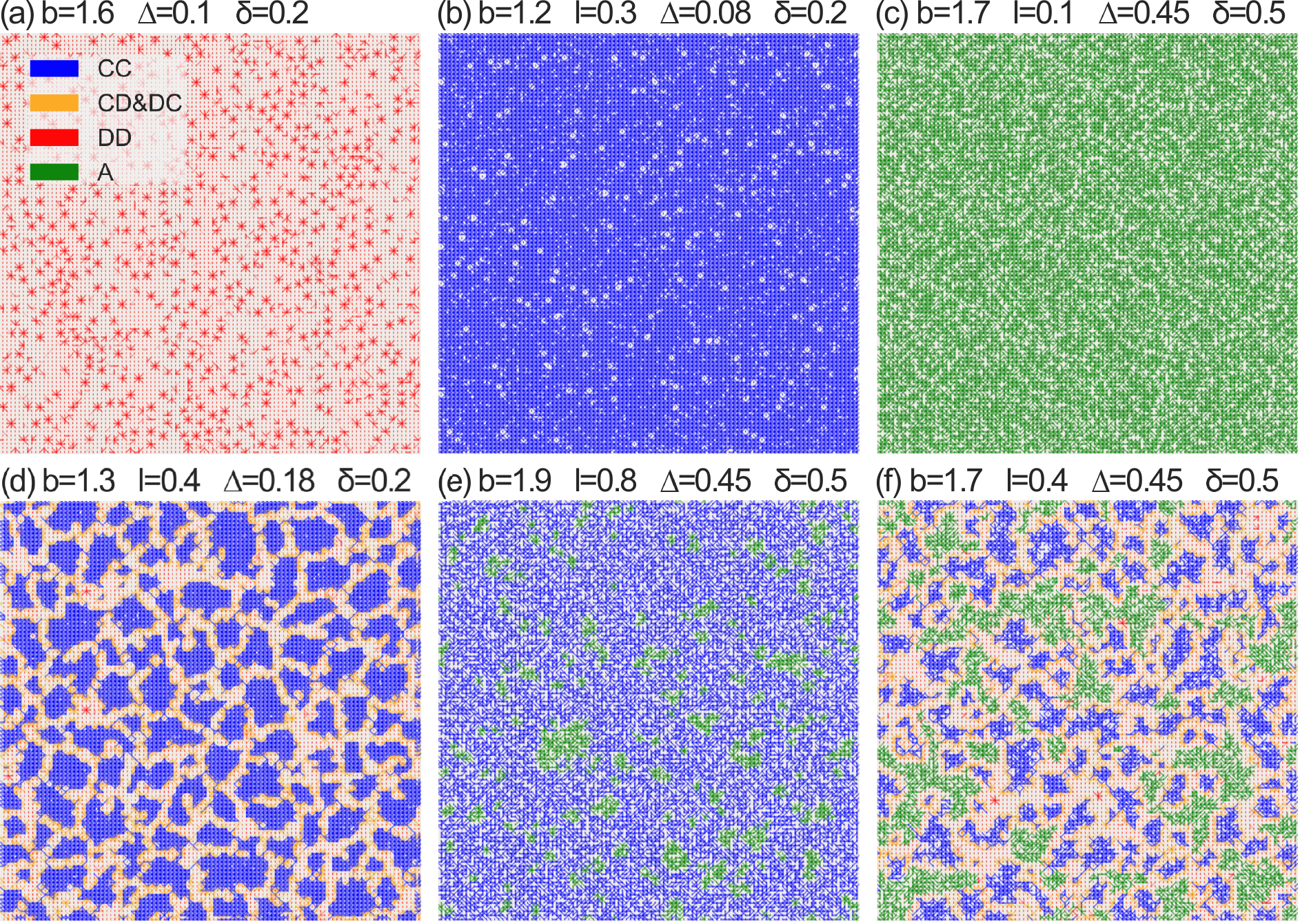, width=0.65\linewidth}}
\caption{
    Typical distribution of edges after a sufficiently long thermalization time
    ($10^6$ MCS).  The opacity of each edge represents its current weight ($w$)
    in the network and is adjusted to be transparent at minimum (i.e.,
    ${w=1-\delta}$) and bright at maximum (i.e., $w=1+\delta$).  The pattern
    \textbf{a} is only possible for the CPD game, which can also exhibit the
    patterns \textbf{b} and \textbf{d}. The patterns \textbf{b-f} can be
    observed in the COPD game.
}
\label{fig:typical_dist}
\end{figure}

Figure~\ref{fig:typical_dist} shows the typical distributions of the pairs of
strategies (i.e., the edges) in the stationary or quasi-stationary states.
In these snapshots we use colors to differentiate the types of edges and
opacity to differentiate the link weights, where, for each scenario,
a transparent edge means that the weight is at minimum (i.e., $w=1-\delta$)
and a bright edge means that the weight is at maximum (i.e., $w=1+\delta$).
As expected, considering that $l>0$,
the pattern \textit{all D} (Fig.~\ref{fig:typical_dist}a) is only possible in
the CPD game. Moreover, for the CPD game, it is also possible to observe
the patterns \textit{all C} (Fig.~\ref{fig:typical_dist}b) and
\textit{C+D} (Fig.~\ref{fig:typical_dist}d).
For the COPD game, all other patterns are also possible, i.e.,
\textit{all A} (Fig.~\ref{fig:typical_dist}c),
\textit{C+A} (Fig.~\ref{fig:typical_dist}e) and
\textit{C+D+A} (Fig.~\ref{fig:typical_dist}f) phases can also be observed.
Of course, the size of the clusters and the average link weight at the stationary
state will depend on the parameter settings. However, for any scenario, it was
observed that the population always evolves to one of these patterns.
Further analysis on the effects of varying the parameter settings have been
shown in previous studies \cite{CardinotSCI,CardinotAICS,Huang2015}.

\begin{figure}[t]
\centering
{\epsfig{file=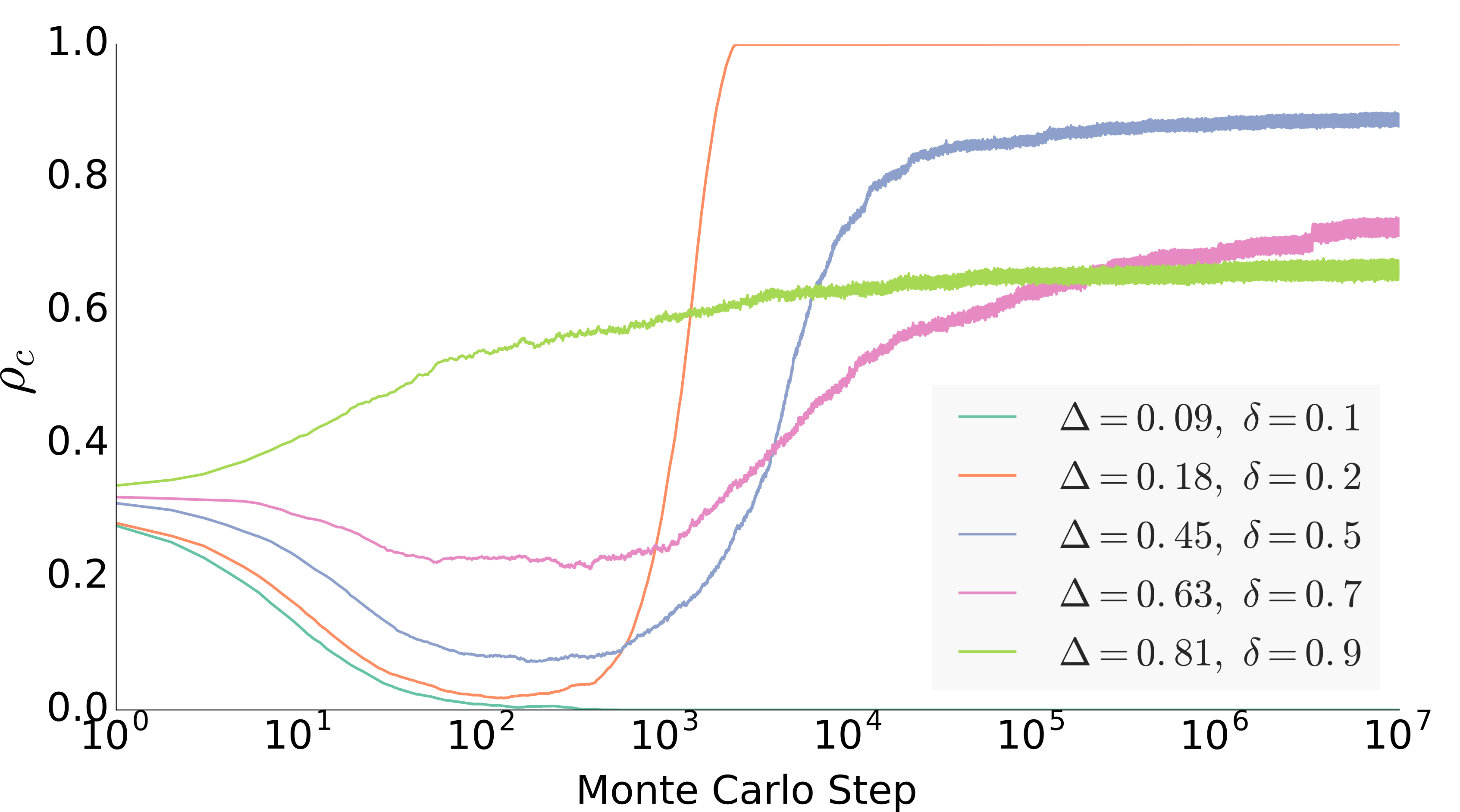, width=0.50\linewidth}}
\caption{
    Time course of the fraction of cooperation for different values of $\Delta$
    and $\delta$ when $b=1.9$, $l=0.8$ and $\Delta/\delta=0.9$.  Contrary to
    what Fig.~\ref{fig:ternary} may suggest, here we see that high $\delta$ is
    not always the best option to promote cooperation.
}
\label{fig:convergence}
\end{figure}

Another interesting result is that, although previous research has claimed that
``intermediate link weight amplitude can provide best environment for the
evolution of cooperation'' \cite{Huang2015}, our experiments reveal that there
is no global optimal value of $\Delta/\delta$ (defined by Huang et al. \cite{Huang2015} as the link weight
amplitude) nor $\delta$ for all environmental settings. Moreover, despite the
fact that high $\delta$ usually leads to more cooperation, it does not mean
that high $\delta$ is always the best option. Fig.~\ref{fig:convergence}, for
example, illustrates a scenario in which high $\delta$ is actually a bad choice.
In fact, as already expected (Section~\ref{sec:methods}), in many cases
it is possible to observe that the population evokes the existence of
Griffiths-like phases, which makes it very difficult for the system to converge to
a stationary state. For instance, the population evolves into a frozen pattern
in the scenarios shown in Fig.~\ref{fig:typical_dist}d and Fig.~\ref{fig:typical_dist}e;
moreover, the curves for $\{\Delta=0.45,\delta=0.5\}$, $\{\Delta=0.63,\delta=0.7\}$ and
$\{\Delta=0.81,\delta=0.9\}$ in Fig.~\ref{fig:convergence} are also evidence of the
same technical difficulties in classifying the stationary state. Note that, in some
scenarios, the presence of the cyclic dominance for the COPD (i.e., coexistence of
the three strategies as observed in Fig.~\ref{fig:typical_dist}f) may eliminate
the emergence of the frozen patterns when the population has only two
strategies, i.e., C+D or C+A. This phenomenon has been discussed in the literature
of evolutionary games \cite{Perc2017,Perc2011,Droz2009}. Furthermore, the
dynamical behaviour observed in Fig.~\ref{fig:convergence} illustrates the nature
of enhanced network reciprocity \cite{Nowak2006b} promoted when $\delta>0.1$.
In these scenarios, we can see that defectors are dominated by abstainers, allowing
a few clusters of cooperators to survive; as a result of the absence of
defectors, cooperators invade most (or all) of the abstainers in the population,
which explains the initial drop, and the subsequent recovery, of the fraction of
cooperators in Fig.~\ref{fig:convergence}. Similar behaviour has also been
observed in previous work \cite{Szolnoki2008a}.

These results motivate the search for a better understanding of the
evolutionary dynamics of the link weights. In the following sections, we
will discuss how the link weights evolve over time.

\subsection{Understanding how the link weights evolve}
\label{sec:barchart}

In order to better understand how the link weights between agents evolve over
time, we investigate the distribution of link weights for different
values of $b$ (temptation to defect), $\Delta$ and $\delta$, for both the
Coevolutionary Prisoner's Dilemma (CPD) \cite{Huang2015} and the Coevolutionary Optional
Prisoner's Dilemma (COPD) \cite{CardinotSCI} games, where the latter also
involves the variation of the loner's payoff ($l$).

\begin{figure}[t]
\centering
{\epsfig{file=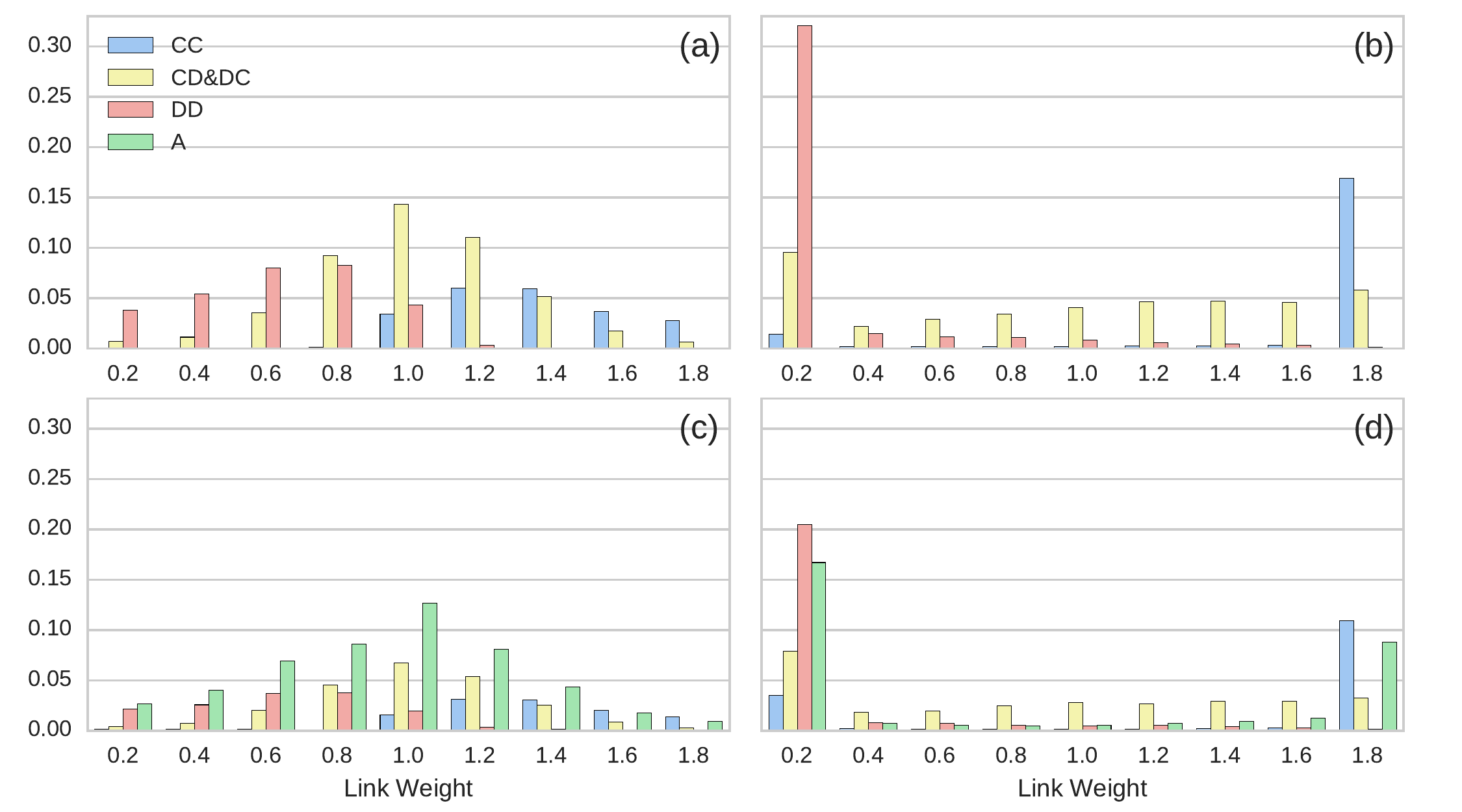, width=0.85\linewidth}}
\caption{
    Distribution of link weight at the Monte Carlo steps 1 (left) and 5 (right)
    for a population of $102 \times 102$ agents playing the Coevolutionary Prisoner's
    Dilemma game (top) and the Coevolutionary Optional Prisoner's Dilemma game (bottom)
    for $l=0.2$, $b=1.6$, $\Delta=0.2$ and $\delta=0.8$. It was observed that the
    initial behaviour of all types of links is the same for both games regardless of the
    ratio $\Delta/\delta$, i.e., the initial link weight distribution looks the same
    for all scenarios, where the only difference is in the proportion of each type,
    which depends on the values of all parameter settings.
}
\label{fig:barchart}
\end{figure}

Figure~\ref{fig:barchart} shows the distribution of link weights for each type
of agent interaction when $b=1.6$, $l=0.2$, $\Delta=0.2$ and $\delta=0.8$ which
is representative of the outcomes of other values as well.
As discussed previously, we know that the ratio $\Delta/\delta$ can be used to
determine the number of link weights that an agent is allowed to have, which
is actually evidenced when the link weight distribution is plotted over time.
Despite the fact that the percentage of each type of link varies according to
factors such as the total number of states and the value of $b$ and $l$ (for the COPD
game), which will consequently affect the final outcome, it was observed (as shown in
Figure~\ref{fig:barchart}) that for any ratio $\Delta/\delta$ the initial dynamics of
all types of links is exactly the same for both games, that is:
\begin{itemize}

    \item \textbf{Observation 1} Defector-Defector (DD) tends to move to states of
        lowest link weight.

    \item \textbf{Observation 2} Cooperator-Defector (CD) and Defector-Cooperator (DC)
        move to the extremes, keeping a small amount of intermediate states.

    \item \textbf{Observation 3} Cooperator-Cooperator (CC) tends to move to states of
        highest link weight, but will also occupy the state of lowest link
        weight as the DCs and ACs will eventually become CCs.

    \item \textbf{Observation 4} Abstainers (AC, AD, CA, DA or AA) move to the extremes.

\end{itemize}

Considering that utility is obtained by the product of link weight and payoff
(Eq.~\ref{eq:utility}), and that the payoff of DD and CD is equal to zero
(Eq.~\ref{eq:payoff}), the utility ($u_{xy}$) associated with these link types
will always be equal to zero, which is always the worst case as $u_{xy} \ge 0.0$.
In this way, these agents will always be punished by $\Delta$
(Eq.~\ref{eq:bigdelta}) and consequently, occupy states of lowest link
weight (Observation 1).

Also, note that CD and DC are unstable configurations as the first will always
get $u_{xy}=0.0$ and the second is prone to get higher utilities as the
temptation to defect ($b$) is always the highest payoff
(Eq.~\ref{eq:payoff}). Thus, these agents are constantly receiving
${\pm \Delta}$, which explains the phenomenon of having a small number of them
along the intermediate states (Observation 2).

Although the payoff obtained by mutual cooperators (CC) is smaller than the one
obtained by a defector-cooperator interaction (DC), i.e., $T > R$, the mutual
cooperators are much more stable than DCs as both agents always get the same
payoff (i.e., $R$). That is exactly the reason why these agents tend to a maximum
link weight (Observation 3). Also, note that as the link weight is updated based on
the comparison of the local utility of each connection with the average utility
of the eight neighbours, when a cluster of nine cooperators is formed (i.e.,
one cooperator surrounded by eight cooperators), their links will remain in
equilibrium, where the average link weight will tend to the value of $R$.
In this way, for most scenarios of full dominance of cooperation, approximately
half of the links will have a minimum weight and the other half will have a
maximum weight.

It is noteworthy that we count all types of Abstainer's connections (AC, CA,
AD, DA and AA) together because in the Optional Prisoner's Dilemma game, when a
agent abstains, both agents receive the same payoff ($l$). In this way, the
main reason why abstainers move to the extremes in Figure~\ref{fig:barchart} is
that DD and CD agents ($u_{xy}=0.0$) tend to abstain to increase their local
utility ($u_{xy}>0.0$) becoming ADs, which consequently are allocated in states
of lowest link weight. For the same reason, abstention might be the best option
in mixed clusters of C's and D's, where the chances of getting $u_{xy}=0.0$
increases, then the agents may tend to abstain, eventually going to states
of highest link weight (Observation 4).

Moreover, we point out that, as we force all link weights to be within the range
$1-\delta$ to $1+\delta$, the phenomenon of having more agents occupying the
maximum and minimum states is clearly expected. However, the observation of
the initial dynamics of both games being the same for any combination of the
parameters (i.e., $b$, $l$, $\Delta$ and $\delta$) is a counter-intuitive
result, which in turn shows that the observations discussed above are valid for
both models.

\subsection{Investigating the role of heterogeneity}

The reason why higher values of $\delta$ promote cooperation best remains one
of the central open questions in this model.  Based on the results discussed in
previous sections, we know that the link weights usually evolve
heterogeneously, which makes the effective payoff matrix unpredictable, adding
a new layer of complexity to the model. For instance, in the traditional
Prisoner's Dilemma game, any defector who plays with a cooperator will always
get the value of the constant $b$; however in the coevolutionary model, this is
unpredictable and heterogeneous as each defector-cooperator interaction might
be in a different state.

\begin{figure}[t]
\centering
{\epsfig{file=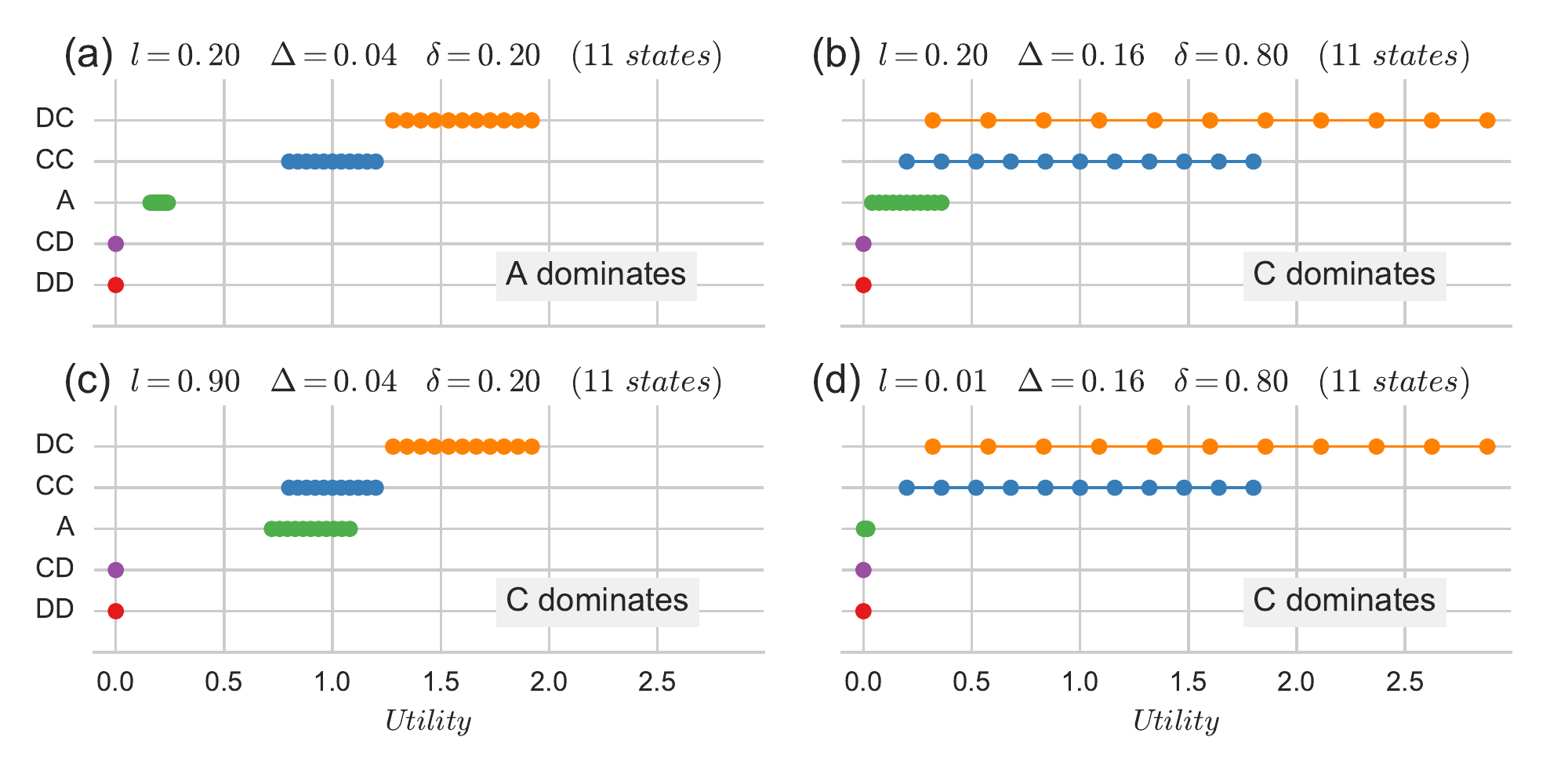, width=0.8\linewidth}}
\caption{
    Possible utilities (i.e., the product of payoff and link weight) for each type
    of edge, i.e., \textit{Defector-Cooperator} (DC), \textit{Cooperator-Cooperator} (CC),
    \textit{Cooperator-Defector} (CD), \textit{Defector-Defector} (DD) and the ones with one
    or two \textit{Abstainers} (A). We observe that there is a connection between the overlap
    of possible utilities and the final outcome.
}
\label{fig:utility}
\end{figure}

Considering that the boundary states and the set of possible link
weights is determined by the parameters $\Delta$ and $\delta$, we can calculate
all the possible utilities for each type of edge (i.e., CD, DC, DD, CC and A),
which may allow us to better understand how the parameter settings affect the
interplay between the evolution of strategies and their possible utilities.  In
this way, Figure~\ref{fig:utility} shows the shape of all possible utilities
for four different scenarios, all for the same temptation to defect ($b=1.6$)
and the same number of states ($\Delta/\delta=0.2$, i.e., $11$ states). Monte
Carlo simulations revealed that cooperation is the dominant strategy in the
scenarios of Fig.~\ref{fig:utility}b-d; and that abstention dominates in
Fig.~\ref{fig:utility}a.

In fact, when we plot the possible utilities side by side
(Fig.~\ref{fig:utility}) we can see that the outcomes obtained through Monte
Carlo simulations were actually expected. For instance, in
Fig.~\ref{fig:utility}a, the DC connections will always be the most
profitable option in the initial steps, which in turn make the population of
cooperators die off. After that, with the lack of cooperators in the
population, DC is not possible anymore and abstention starts to be the best
option as its payoff is always greater than the punishment for mutual defection
(i.e., $lw_a>0$).

However, notice that when the value of $\delta$ is increased (i.e.,
Fig.~\ref{fig:utility}b),  the overlap between the possible utilities for
each type of connection also increases. In this case, we see that DC is
the best option only in $5/11$ of the cases, which enable cooperators to
survive and as DC tends to minimum, CC will tend to maximum and abstention is
sometimes better than DC (Section~\ref{sec:barchart}). Thus, the dominance of
cooperators is also expected. We can also observe that due to the huge overlap
of DCs and CCs, even if the loner's payoff is very low, i.e.,
Fig.~\ref{fig:utility}d, or if abstention does not exist (i.e., CPD game),
cooperation would still be expected.

Moreover, Fig.~\ref{fig:utility}c illustrates interesting evidence of how abstention
can support cooperation. The only difference between this scenario and
Fig.~\ref{fig:utility}a is the value of $l$. At a first glance, intuition may
lead us to believe that if Fig.~\ref{fig:utility}a with $l=0.2$ resulted in
full dominance of abstainers, increasing the value of $l$ would just make the
option to abstain more profitable, which consequently would not change the
outcome.  Surprisingly, this does not occur. Actually, abstainers only dominate
the whole population when the population of cooperators is decimated. In this
way, despite the fact that DC is still the best option and that the population
of cooperators tend to decrease, they will not die off. Then, when the
population of defectors become too high, they will prefer to turn into
abstainers and with the increase of abstention, mutual cooperation will now be
the best option, which allows abstainers to fully dominate the environment.

Thus, results show that if any of the possible DC, CC and A utilities do not
overlap, then abstention will be the dominant strategy for the COPD game and
defection will be the dominant strategy for the CPD game, except of course,
when $b$ is too low (i.e. $b<1.1$), which usually promotes the coexistence of
the available strategies. In general, we can observe that the greater the
overlap between DC and CC utilities or/and CC and A utilities, then the more
chances cooperators have to survive and dominate.

Notice that both the loner's payoff ($l$) and the link weights (which are
controlled by the parameters $\Delta$ and $\delta$) are actually mechanisms to
weaken the benefits of defecting (i.e., effective utility of DC). As the
parameter $\delta$ will act in the expansion of the utility boundaries, the
greater the value of $\delta$, the greater are the number of cases in which CC
overlaps DC, which in turn promote cooperation best. The same scenario occurs
when CC and A overlap, which will work as an extra mechanism to strengthen
cooperators. That is, when the overlap of DC and CC is scarce or absent,
overlapping CC and A can help cooperators to survive. This also explains why
COPD is better than CPD in adverse scenarios~\cite{CardinotSCI}.

The drawback of a large overlap of utilities is that the population may evolve
into a frozen pattern in which a Griffiths-like phase can occur. In these
scenarios, it might be very difficult to reach the full dominance of cooperative
behaviour (Fig.~\ref{fig:convergence}). Thus, higher values of $\delta$ may
promote cooperation best in a wider range of scenarios, but it might evoke the
presence of frozen patterns of C+A. It is noteworthy that as the utility overlaps
are also dependent of the values of $b$ and $l$, all parameter settings
may, in fact, influence the emergence of these frozen patterns.

\section{Conclusions}
\label{sec:conclusion}

This work investigates the role of heterogeneity in a population of agents
playing the Prisoner's Dilemma (PD) game and the Optional Prisoner's Dilemma
(OPD) game on a weighted square network with boundary conditions.
Coevolutionary rules are adopted, enabling both the game strategies and the
network to evolve over time, leading to the
so-called Coevolutionary Prisoner's Dilemma (CPD) and the Coevolutionary Optional
Prisoner's Dilemma (COPD) games respectively.
A number of Monte Carlo simulations are performed in which each agent is
initially assigned to a strategy with equal probability (i.e., random
initial distribution of strategies).
Echoing the findings of previous research~\cite{CardinotSCI}, we show that
independently of the link weight heterogeneity, the COPD game is still much more
beneficial for the emergence of cooperation than the traditional OPD or the CPD
games. Moreover, although previous research has claimed the
opposite~\cite{Huang2015}, we show that there is no global optimal value of the
parameters $\Delta$ and $\delta$ for all environmental settings.

Experiments revealed that the correlation between the emergence of cooperation
and heterogeneity does not hold for all scenarios, indicating that
heterogeneity itself does not favour cooperation. Actually, it was observed
that the higher the heterogeneity of states, the greater the chance of
overlapping states, which is the actual mechanism for promoting cooperation.
Namely, when considering the COPD game, if any of the possible
\textit{Defector-Cooperator} (DC), \textit{Cooperator-Cooperator} (CC) and
\textit{Abstention} (A) utilities do not overlap, then abstainers dominate the
environment; while for the CPD game, defection will be the dominant strategy.
In general, we observed that the greater the overlap between DC and CC utilities
or/and CC and A utilities, the more chances cooperators have to survive and dominate.

Finally, we highlight that both the loner's payoff and the link weights are
actually mechanisms that weaken the benefits of defecting. In addition, abstention
also works as an extra mechanism to strengthen cooperators, which explains why
COPD is better than CPD in adverse scenarios.
We believe that it might be possible to analytically define, through the analysis
of utility overlap, which is the best value of $\delta$ for a given payoff matrix.
Also, considering this model for regular graphs, it might be interesting to
consider pair approximation techniques to describe the evolutionary dynamics of
weighted networks \cite{Baalen2000,Ohtsuki2006}.  To conclude, this paper
provides a novel perspective for understanding cooperative behaviour in a
dynamic network, which resembles a wide range of real-world scenarios. We hope
this paper can serve as a basis for further research on the role of utility
overlap to advance the understanding of the evolution of cooperation in
coevolutionary spatial games.

\section*{Acknowledgement}
This work was supported by the National Council for Scientific and
Technological Development (CNPq-Brazil). Grant number 234913/2014-2.

\section*{References}
\bibliographystyle{elsarticle-num}
\bibliography{refs}

\begin{thebibliography}{10}
\expandafter\ifx\csname url\endcsname\relax
  \def\url#1{\texttt{#1}}\fi
\expandafter\ifx\csname urlprefix\endcsname\relax\def\urlprefix{URL }\fi
\expandafter\ifx\csname href\endcsname\relax
  \def\href#1#2{#2} \def\path#1{#1}\fi

\bibitem{Nowak2006}
M.~A. Nowak, Evolutionary dynamics, Harvard University Press, 2006.

\bibitem{Smith1982}
J.~M. Smith, Evolution and the Theory of Games, Cambridge University Press,
  1982.

\bibitem{Perc2017}
M.~Perc, J.~J. Jordan, D.~G. Rand, Z.~Wang, S.~Boccaletti, A.~Szolnoki,
  Statistical physics of human cooperation, Physics Reports 687 (2017) 1--51.

\bibitem{Nowak2009}
M.~A. Nowak, C.~E. Tarnita, T.~Antal, Evolutionary dynamics in structured
  populations, Philosophical Transactions of the Royal Society of London B:
  Biological Sciences 365~(1537) (2009) 19--30.

\bibitem{Lieberman2005}
E.~Lieberman, C.~Hauert, M.~A. Nowak, Evolutionary dynamics on graphs, Nature
  433~(7023) (2005) 312--316.

\bibitem{Nowak1992}
M.~A. Nowak, R.~M. May, Evolutionary games and spatial chaos, Nature 359~(6398)
  (1992) 826--829.

\bibitem{Szolnoki2016}
A.~Szolnoki, M.~Perc, Leaders should not be conformists in evolutionary social
  dilemmas, Scientific Reports 6 (2016) 23633.

\bibitem{Xia2015}
C.-Y. Xia, S.~Meloni, M.~Perc, Y.~Moreno, Dynamic instability of cooperation
  due to diverse activity patterns in evolutionary social dilemmas, EPL 109~(5)
  (2015) 58002.

\bibitem{Santos2005}
F.~C. Santos, J.~M. Pacheco, Scale-free networks provide a unifying framework
  for the emergence of cooperation, Phys. Rev. Lett. 95 (2005) 098104.

\bibitem{Chen2008}
X.~Chen, L.~Wang, Promotion of cooperation induced by appropriate payoff
  aspirations in a small-world networked game, Physical Review E 77 (2008)
  017103.

\bibitem{Fu2007}
F.~Fu, L.-H. Liu, L.~Wang, Evolutionary prisoner's dilemma on heterogeneous
  newman-watts small-world network, The European Physical Journal B 56~(4)
  (2007) 367--372.

\bibitem{Altrock2017}
P.~M. Altrock, A.~Traulsen, M.~A. Nowak, Evolutionary games on cycles with
  strong selection, Phys. Rev. E 95 (2017) 022407.

\bibitem{Teng2014}
X.~Teng, S.~Yan, S.~Tang, S.~Pei, W.~Li, Z.~Zheng, Individual behavior and
  social wealth in the spatial public goods game, Physica A: Statistical
  Mechanics and its Applications 402 (2014) 141--149.

\bibitem{Pena2012}
J.~Peña, Y.~Rochat, Bipartite graphs as models of population structures in
  evolutionary multiplayer games, PLOS ONE 7~(9) (2012) 1--13.

\bibitem{Gomez2011}
J.~Gómez-Gardeñes, M.~Romance, R.~Criado, D.~Vilone, A.~Sánchez,
  Evolutionary games defined at the network mesoscale: The public goods game,
  Chaos: An Interdisciplinary Journal of Nonlinear Science 21~(1) (2011)
  016113.

\bibitem{Szabo2007}
G.~Szabó, G.~Fáth, Evolutionary games on graphs, Physics Reports 446~(4)
  (2007) 97--216.

\bibitem{Perc2013}
M.~Perc, J.~G{\'o}mez-Garde{\~n}es, A.~Szolnoki, L.~M. Flor{\'\i}a, Y.~Moreno,
  Evolutionary dynamics of group interactions on structured populations: a
  review, Journal of The Royal Society Interface 10~(80) (2013) 20120997.

\bibitem{Zhang2011a}
C.~Y. Zhang, J.~L. Zhang, G.~M. Xie, L.~Wang, Coevolving agent strategies and
  network topology for the public goods games, The European Physical Journal B
  80~(2) (2011) 217--222.

\bibitem{Zhang2011b}
C.~Zhang, J.~Zhang, G.~Xie, L.~Wang, M.~Perc, Evolution of interactions and
  cooperation in the spatial prisoner's dilemma game, PLOS ONE 6~(10) (2011)
  1--7.

\bibitem{Hai2011}
H.-F. Zhang, R.-R. Liu, Z.~Wang, H.-X. Yang, B.-H. Wang, Aspiration-induced
  reconnection in spatial public-goods game, EPL (Europhysics Letters) 94~(1)
  (2011) 18006.

\bibitem{Szolnoki2009}
A.~Szolnoki, M.~Perc, Resolving social dilemmas on evolving random networks,
  EPL (Europhysics Letters) 86~(3) (2009) 30007.

\bibitem{Zimmermann2004}
M.~G. Zimmermann, V.~M. Egu\'{\i}luz, M.~San~Miguel, Coevolution of dynamical
  states and interactions in dynamic networks, Physical Review E 69 (2004)
  065102.

\bibitem{Zimmermann2001}
M.~G. Zimmermann, V.~M. Egu{\'i}luz, M.~S. Miguel, {Cooperation, Adaptation and
  the Emergence of Leadership}, Springer, Berlin, Heidelberg, 2001, pp. 73--86.

\bibitem{Perc2010}
M.~Perc, A.~Szolnoki, {Coevolutionary games -- A mini review}, Biosystems
  99~(2) (2010) 109--125.

\bibitem{Rapoport1965}
A.~Rapoport, A.~M. Chammah, Prisoner's dilemma: A study in conflict and
  cooperation, Vol. 165, University of Michigan Press, 1965.

\bibitem{Szabo2002}
G.~Szab\'o, C.~Hauert, Evolutionary prisoner's dilemma games with voluntary
  participation, Physical Review E 66 (2002) 062903.

\bibitem{Batali1995}
J.~Batali, P.~Kitcher, Evolution of altruism in optional and compulsory games,
  Journal of Theoretical Biology 175~(2) (1995) 161--171.

\bibitem{Lu2017}
J.~Lu, L.~Wang, Y.-L. Wang, X.~Zhang, Logit selection promotes cooperation in
  voluntary public goods game, Applied Mathematics and Computation 310 (2017)
  134--138.

\bibitem{Hauert2008}
C.~Hauert, A.~Traulsen, H.~Brandt, M.~A. Nowak, {Public goods with punishment
  and abstaining in finite and infinite populations}, Biological Theory 3~(2)
  (2008) 114--122.

\bibitem{CardinotSAB}
M.~Cardinot, M.~Gibbons, C.~O'Riordan, J.~Griffith, Simulation of an optional
  strategy in the prisoner's dilemma in spatial and non-spatial environments,
  in: From Animals to Animats 14 (SAB 2016), Springer International Publishing,
  Cham, 2016, pp. 145--156.

\bibitem{Hauert2003}
C.~Hauert, G.~Szabo, Prisoner's dilemma and public goods games in different
  geometries: compulsory versus voluntary interactions, Complexity 8~(4) (2003)
  31--38.

\bibitem{CardinotSCI}
M.~Cardinot, C.~O'Riordan, J.~Griffith, {The Impact of Coevolution and
  Abstention on the Emergence of Cooperation}, ArXiv e-prints\href
  {http://arxiv.org/abs/1705.00094} {\path{arXiv:1705.00094}}.

\bibitem{CardinotAICS}
M.~Cardinot, J.~Griffith, C.~O’Riordan, {Cyclic Dominance in the Spatial
  Coevolutionary Optional Prisoner’s Dilemma Game}, in: D.~Greene, B.~M.
  Namee, R.~Ross (Eds.), Artificial Intelligence and Cognitive Science 2016,
  Vol. 1751 of CEUR Workshop Proceedings, Dublin, Ireland, 2016, pp. 33--44.

\bibitem{Huang2015}
K.~Huang, X.~Zheng, Z.~Li, Y.~Yang, Understanding cooperative behavior based on
  the coevolution of game strategy and link weight, Scientific Reports 5 (2015)
  14783.

\bibitem{Cao2011}
L.~Cao, H.~Ohtsuki, B.~Wang, K.~Aihara, Evolution of cooperation on adaptively
  weighted networks, Journal of Theoretical Biology 272~(1) (2011) 8 -- 15.

\bibitem{Li2017}
D.~Li, J.~Ma, D.~Han, M.~Sun, L.~Tian, H.~E. Stanley, The co-evolution of
  networks and prisoner’s dilemma game by considering sensitivity and
  visibility, Scientific Reports 7 (2017) 45237.

\bibitem{Fan2017}
R.~Fan, Y.~Zhang, M.~Luo, H.~Zhang, Promotion of cooperation induced by
  heterogeneity of both investment and payoff allocation in spatial public
  goods game, Physica A: Statistical Mechanics and its Applications
  465~(Supplement C) (2017) 454 -- 463.

\bibitem{Iwata2016}
M.~Iwata, E.~Akiyama, Heterogeneity of link weight and the evolution of
  cooperation, Physica A: Statistical Mechanics and its Applications 448 (2016)
  224 -- 234.

\bibitem{Amaral2016}
M.~A. Amaral, L.~Wardil, M.~c.~v. Perc, J.~K.~L. da~Silva, Evolutionary mixed
  games in structured populations: Cooperation and the benefits of
  heterogeneity, Phys. Rev. E 93 (2016) 042304.

\bibitem{Perc2011}
M.~Perc, Does strong heterogeneity promote cooperation by group interactions?,
  New Journal of Physics 13~(12) (2011) 123027.

\bibitem{Wardil2011}
L.~Wardil, J.~K.~L. da~Silva, The evolution of cooperation in heterogeneous
  networks when opponents can be distinguished, Journal of Physics A:
  Mathematical and Theoretical 44~(34) (2011) 345101.

\bibitem{Szolnoki2008a}
A.~Szolnoki, M.~Perc, Z.~Danku, Making new connections towards cooperation in
  the prisoner's dilemma game, EPL (Europhysics Letters) 84~(5) (2008) 50007.

\bibitem{Szolnoki2008b}
A.~Szolnoki, M.~Perc, Coevolution of teaching activity promotes cooperation,
  New Journal of Physics 10~(4) (2008) 043036.

\bibitem{Szolnoki2007}
A.~Szolnoki, G.~Szabó, Cooperation enhanced by inhomogeneous activity of
  teaching for evolutionary prisoner's dilemma games, EPL (Europhysics Letters)
  77~(3) (2007) 30004.

\bibitem{Wang2014}
Z.~Wang, A.~Szolnoki, M.~Perc, Self-organization towards optimally
  interdependent networks by means of coevolution, New Journal of Physics
  16~(3) (2014) 033041.

\bibitem{Griffiths1969}
R.~B. Griffiths, Nonanalytic behavior above the critical point in a random
  ising ferromagnet, Phys. Rev. Lett. 23 (1969) 17--19.

\bibitem{Droz2009}
M.~Droz, J.~Szwabi{\'{n}}ski, G.~Szab{\'o}, Motion of influential players can
  support cooperation in prisoner's dilemma, The European Physical Journal B
  71~(4) (2009) 579.

\bibitem{Nowak2006b}
M.~A. Nowak, Five rules for the evolution of cooperation, Science 314~(5805)
  (2006) 1560--1563.

\bibitem{Baalen2000}
M.~van Baalen, Pair approximations for different spatial geometries, in: L.~R.
  Dieckmann~U, M.~JAJ (Eds.), The Geometry of Ecological Interactions:
  Simplifying Spatial Complexity, Cambridge University Press, 2000, p.
  359–387.

\bibitem{Ohtsuki2006}
H.~Ohtsuki, M.~A. Nowak, The replicator equation on graphs, Journal of
  theoretical biology 243~(1) (2006) 86--97.

\end{thebibliography}

\end{document}